\documentclass[twocolumn,twoside,slac_two]{revtex4}
\usepackage{graphicx}
\usepackage{fancyhdr}
\begin{document}

\title{K-matrix and Dalitz plot analysis from FOCUS}


\author{S. Malvezzi}
\affiliation{I.N.F.N Sezione di Milano Bicocca, Piazza della Scienza 3, 20126
Milano, Italy}
\begin{abstract}
Dalitz analysis is a powerful tool for physics studies within and beyond the
Standard Model. In the last decade it has helped to investigate the Heavy
Flavor hadronic decay dynamics and is now being applied to extract angles of
the CKM Unitarity triangle. To perform such sophisticate analyses we need to
model the strong interaction effects. The FOCUS experiment has performed pilot
studies in the charm sector through the \emph{K-matrix} formalism. What has
been learnt from charm will be beneficial for future accurate beauty
measurements. Experience and results from FOCUS are presented and discussed.
\end{abstract}

\maketitle

\thispagestyle{fancy}


\section{Introduction}
Over the last years we have seen a resurrection of Dalitz plot analyses in
modern Heavy Flavor experiments. This analysis tool, first applied in the
charm sector, has more recently become a standard technique, used for
sophisticated studies and searches for new physics in the beauty sector.
Paradigmatic examples are $B \to \rho \pi$ and $B \to D^{(*)}K^{(*)}$ for the
extraction of the $\alpha$ and $\gamma$ angles of the Unitarity Triangle.
Indeed, the road to go from the detected final states to the intermediate
resonances can be rather insidious and complications arise in both decays.
More precisely, the extraction of $\alpha$ in $B \to \rho \pi$ means,
operatively, selecting and filtering the desired intermediate states among all
the possible $(\pi\pi)\pi$ combinations, e.g. $\sigma \pi$, $f_0(980) \pi$
etc. The extraction of $\gamma$ in $B \to D^{(*)}K^{(*)}$ requires, in turn,
modeling the $D$ amplitudes. This poses the problem of how to deal with
strong-dynamics effects, in particular those regarding the scalar mesons. The
$\pi \pi$ and $K\pi$ S-wave are characterized by broad, overlapping states:
unitarity is not explicitly guaranteed by a simple sum of Breit--Wigner
functions. In addition, independently of the nature of the $\sigma$, it is not
a simple Breit--Wigner. The $f_0(980)$ is a Flatt\'e-like function, and its
lineshape parametrization needs a precise determination of $K K$  and $\pi\pi$
couplings.  Recent analyses of CP violation in the $B \to D K$ channel from
the beauty factories have used the Cabibbo-favored mode $K_s \pi^+\pi^-$,
which is common to both $D^0$ and $\bar D^0$. A set of 16 two-body resonances
had to be introduced to describe the $(K\pi) \pi$ and $K_s (\pi\pi)$ states in
the $D^0$ amplitude: two \emph{ad hoc} resonances were required to reproduce
the excess of events in the $\pi\pi$ spectrum, one at the low-mass threshold,
the other at 1.1\,GeV$^2$. Masses and widths of the two states, named
$\sigma_1$ and $\sigma_2$, were fitted to the data themselves and found to be
$M_{\sigma_1} =484 \pm 9$\,MeV, $\Gamma_{\sigma_1} =383 \pm 14 $ and
$M_{\sigma_2} =1014 \pm 7$\,MeV, $\Gamma_{\sigma_2} =88 \pm 13 $ in BaBar
\cite{gamma_babar} and $M_{\sigma_1} =519 \pm 6$\,MeV, $\Gamma_{\sigma_1} =454
\pm 12 $ $M_{\sigma_2} =1050 \pm 8$\,MeV, $\Gamma_{\sigma_2} =101 \pm 7 $ in
Belle \cite{gamma_belle}. These scalars were invoked with no reference to
those found in other processes, in particular scattering data, and with no
assumption as to the correctness of the physics the model embodies. This
procedure of ``effectively'' fitting data invites a word of caution on
estimating the systematics of these measurements. A question then naturally
arises: in the era of precise measurements, do we know sufficiently well how
to deal with strong-dynamics effects in the analyses?

We have faced parametrization problems in the FOCUS experiment and learnt that
many difficulties are already known and studied in different fields of
physics, such as nuclear and intermediate-energy physics, where broad,
multi-channel, overlapping resonances are treated in the \emph{K-matrix}
formalism. The effort we have had to make mainly consisted in building a
bridge of knowledge and language to reach the high-energy community; our
pioneering work in the charm sector might inspire future accurate studies in
the beauty sector.

\section{The K-matrix and P-vector formalism}

A formalism for studying overlapping and many-channel resonances was proposed
long ago and is based on the  \emph{K-matrix} \cite{wigner,chung}
parametrization. The \emph{K-matrix} formalism provides a direct way of
imposing the two-body unitarity constraint, which is not explicitly guaranteed
in the simple sum of Breit-Wigners, here referred to as the \emph{isobar
model}. Minor unitarity violations are expected for narrow, isolated
resonances but more severe ones exist for broad, overlapping states. This is
the real advantage of the \emph{K-matrix} approach: it heavily simplifies the
formalization of any scattering problem since the unitarity of the S matrix is
automatically encoded.

Originating in the context of two-body scattering, the formalism can be
generalized to cover the case of production of resonances in more complex
reactions \cite{aitch}, with the assumption that the two-body system in the
final state is an isolated one and that the two particles do not
simultaneously interact with the rest of the final state in the production
process \cite{chung}. The validity of the assumed quasi two-body nature of the
process of the \emph{K-matrix} approach can only be verified by a direct
comparison of the model predictions with data. In particular, the failure to
reproduce the Dalitz plot distribution could be an indication of the presence
of relevant, neglected three-body effects.

\section{The FOCUS results}

\subsection{The three pion analysis}

\begin{figure}[h]
\includegraphics[width=70mm]{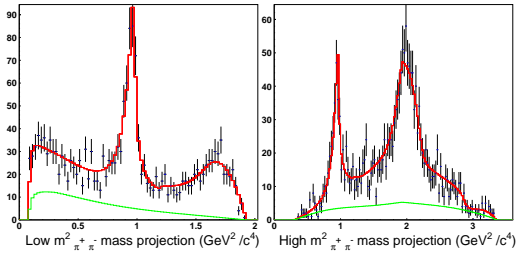}
 \caption{\it
      FOCUS $D_s^+$ Dalitz-plot projections with fit results
superimposed. The background shape under the signal is also shown.
    \label{ds_proj_Kmatrix} }
\end{figure}

\begin{figure}[h]
\includegraphics[width=70mm]{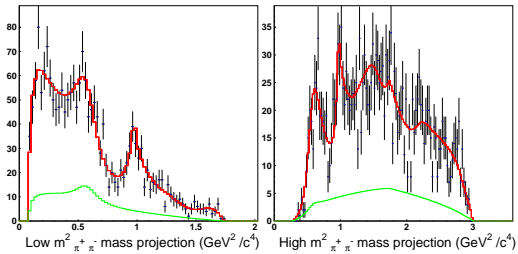}
 \caption{\it
      FOCUS $D^+$ Dalitz-plot projections with fit results
superimposed. The background shape under the signal is also shown.
    \label{dp_proj_Kmatrix} }
\end{figure}

The FOCUS collaboration has implemented the \emph{K-matrix} approach in the
$D_s$ and $D^+ \to \pi^+\pi^-\pi^+$ analyses. It was the first application of
this formalism in the charm sector. Results and details can be found in
\cite{Focus_kmat}. Here I only reprodece plots of the final results. In
Fig.~\ref{ds_proj_Kmatrix} and Fig.~\ref{dp_proj_Kmatrix} the Dalitz-plot
projections are shown for $D_s$ and $D^+$ into three pions.

\begin{figure}[h]
\includegraphics[width=50mm]{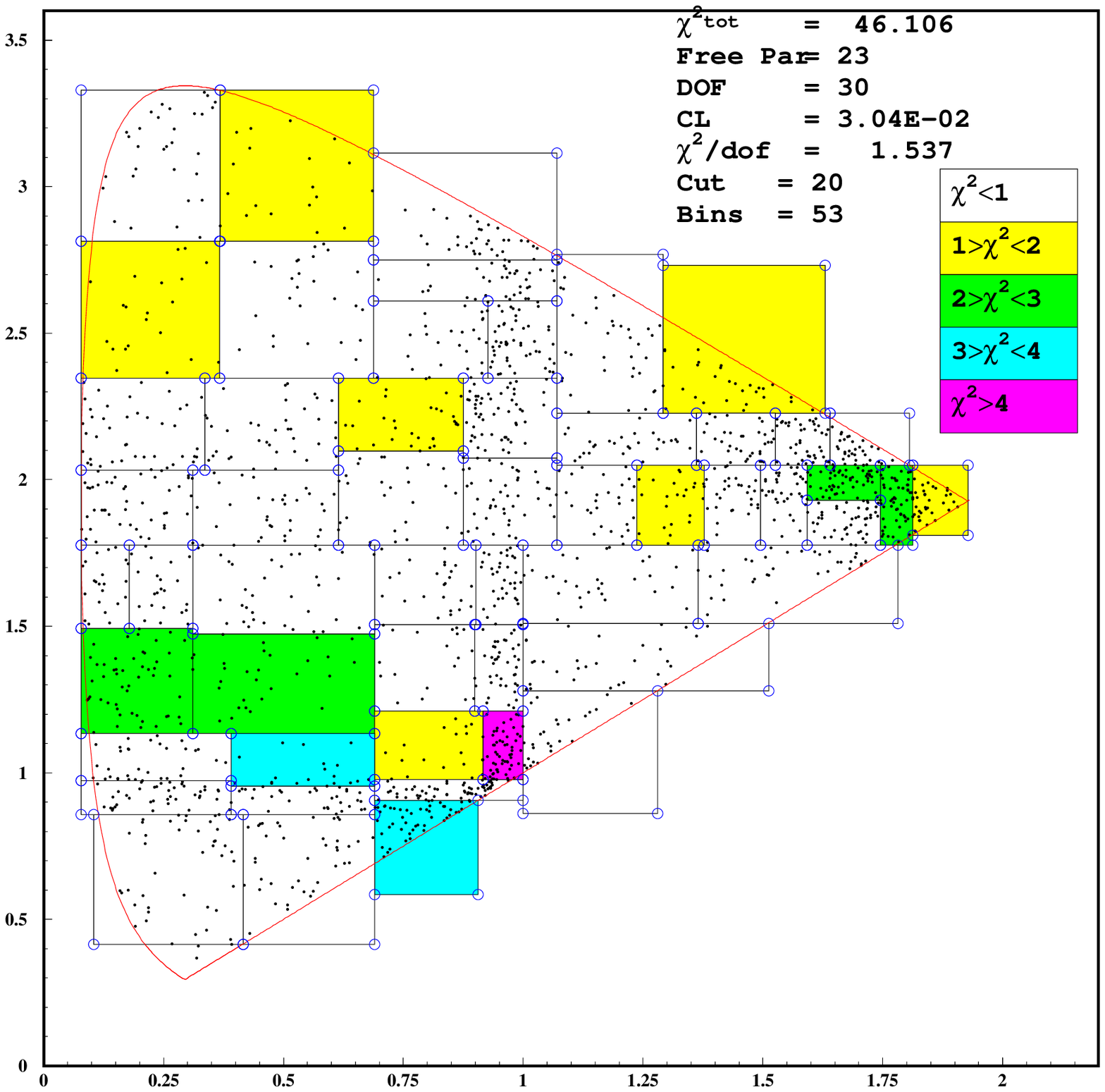}
\includegraphics[width=50mm]{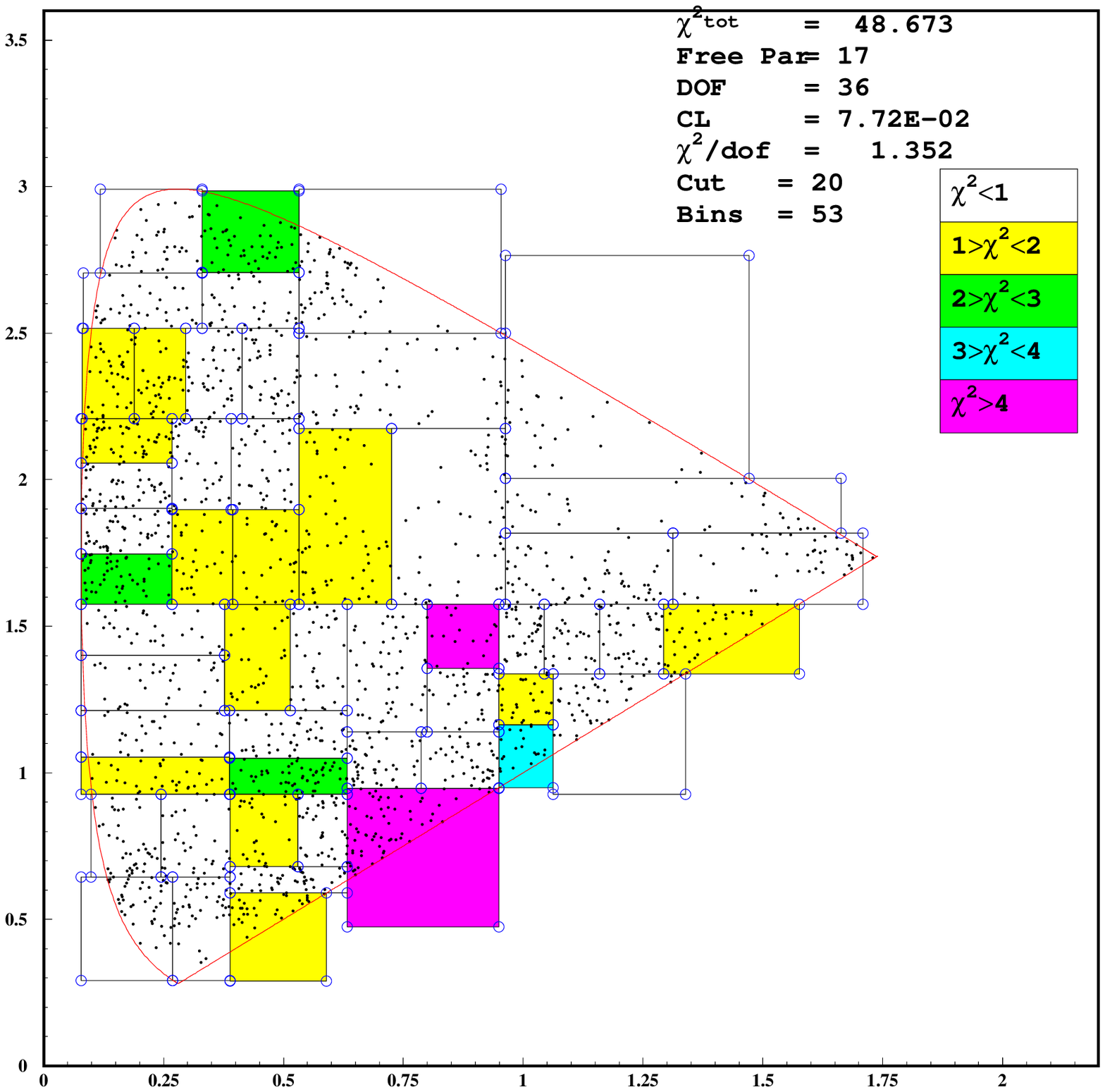}
\ \caption{\it
 $D_s$ and $D^+$ adaptive binning Dalitz-plots for the three pion FOCUS K-matrix
fit.
    \label{d_adapt_km} }
\end{figure}

In Fig.~\ref{d_adapt_km} the FOCUS adaptive binning schemes for $D^+_s$ and
$D^+$ are plotted. In this model \cite{aitch}, the production process, i.e,
the D decay, can be viewed as consisting of an initial preparation of states,
described by the \emph{P-vector}, which then propagates according to
\mbox{$(I-iK\rho)^{-1}$} into the final one. The \emph{K-matrix} here is the
scattering matrix and is used as fixed input in our analysis. Its form was
inferred by the global fit to a rich set of data performed in \cite{anisar1}.
It is interesting to note that this formalism, beside restoring the proper
dynamical features of the resonances, allows for the inclusion in $D$ decays
of the knowledge coming from scattering experiments, i.e, an enormous amount
of results and science. No re-tuning of the \emph{K-matrix} parameters was
needed. The confidence levels of the final fits are 3.0\,\%  and 7.7\,\% for
the $D_s$ and $D^+$ respectively. The results were extremely encouraging since
the same \emph{K-matrix} description gave a coherent picture of both two-body
scattering measurements in light-quark experiments \emph{as well as}
charm-meson decay. This result was not obvious beforehand. Furthermore, the
same model was able to reproduce features of the $D^+\to\pi^+\pi^-\pi^+$
Dalitz plot that would otherwise require an \emph{ad hoc} $\sigma$ resonance.
The better treatment of the $S$-wave contribution provided by the
\emph{K-matrix} model was able reproduce the low-mass $\pi^+\pi^-$ structure
of the $D^+$ Dalitz plot. This suggests that any $\sigma$-like object in the
$D$ decay should be consistent with the same $\sigma$-like object measured in
$\pi^+\pi^-$ scattering.

Further considerations and conclusions from the FOCUS three-pion analysis were
limited by the sample statistics, i.e. $1475 \pm 50$ and $1527 \pm 51$ events
for $D_s$ and $D^+$ respectively. We considered mandatory to test the
formalism at higher statistics. This was accomplished by the $D^+ \to
K^-\pi^+\pi^+$ analysis.

\subsection{The \boldmath$D^+ \to K^-\pi^+\pi^+$}

\begin{figure}[!ht] \centering
  {
 \includegraphics[width=40mm]{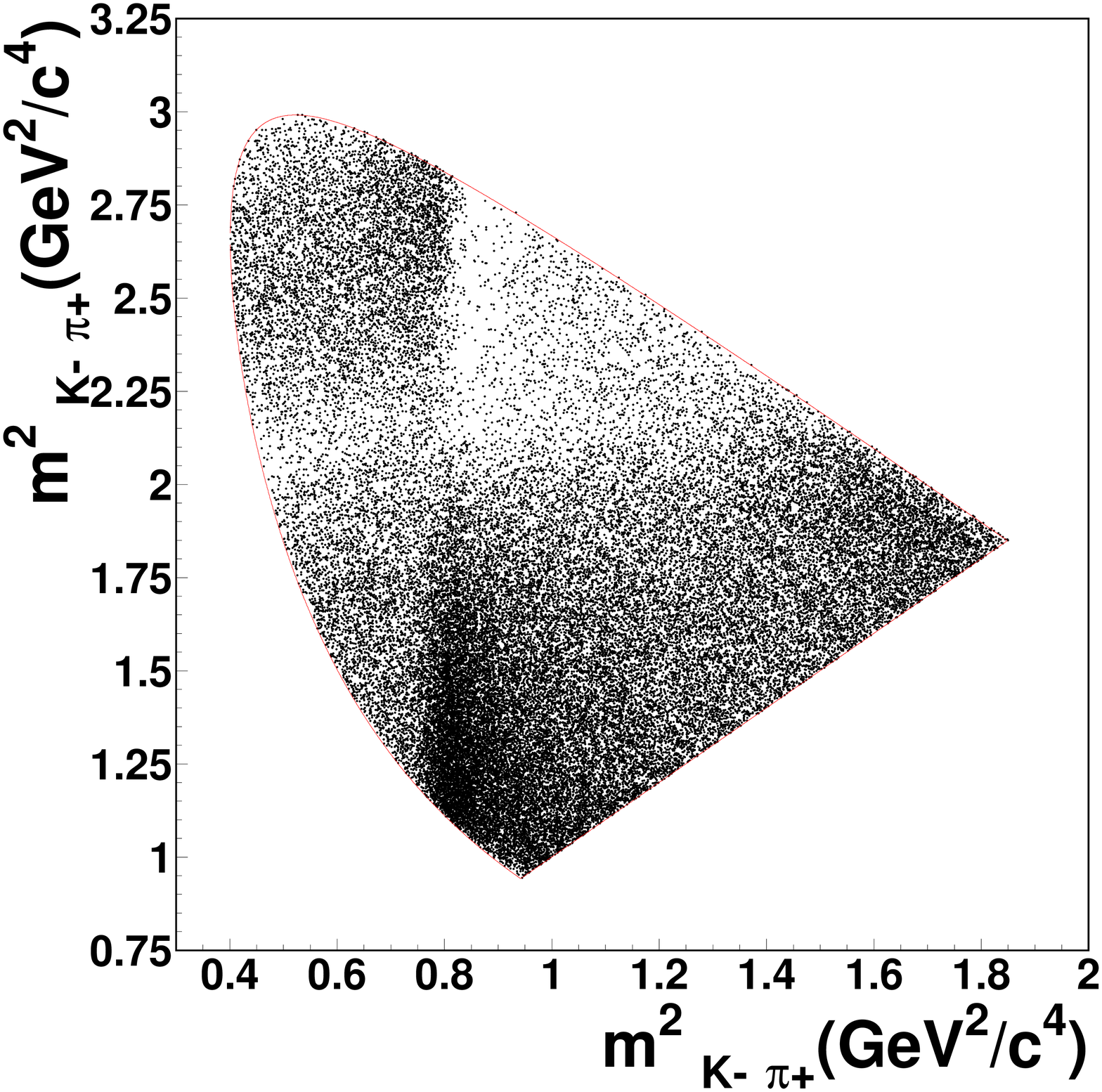}
 \includegraphics[width=40mm]{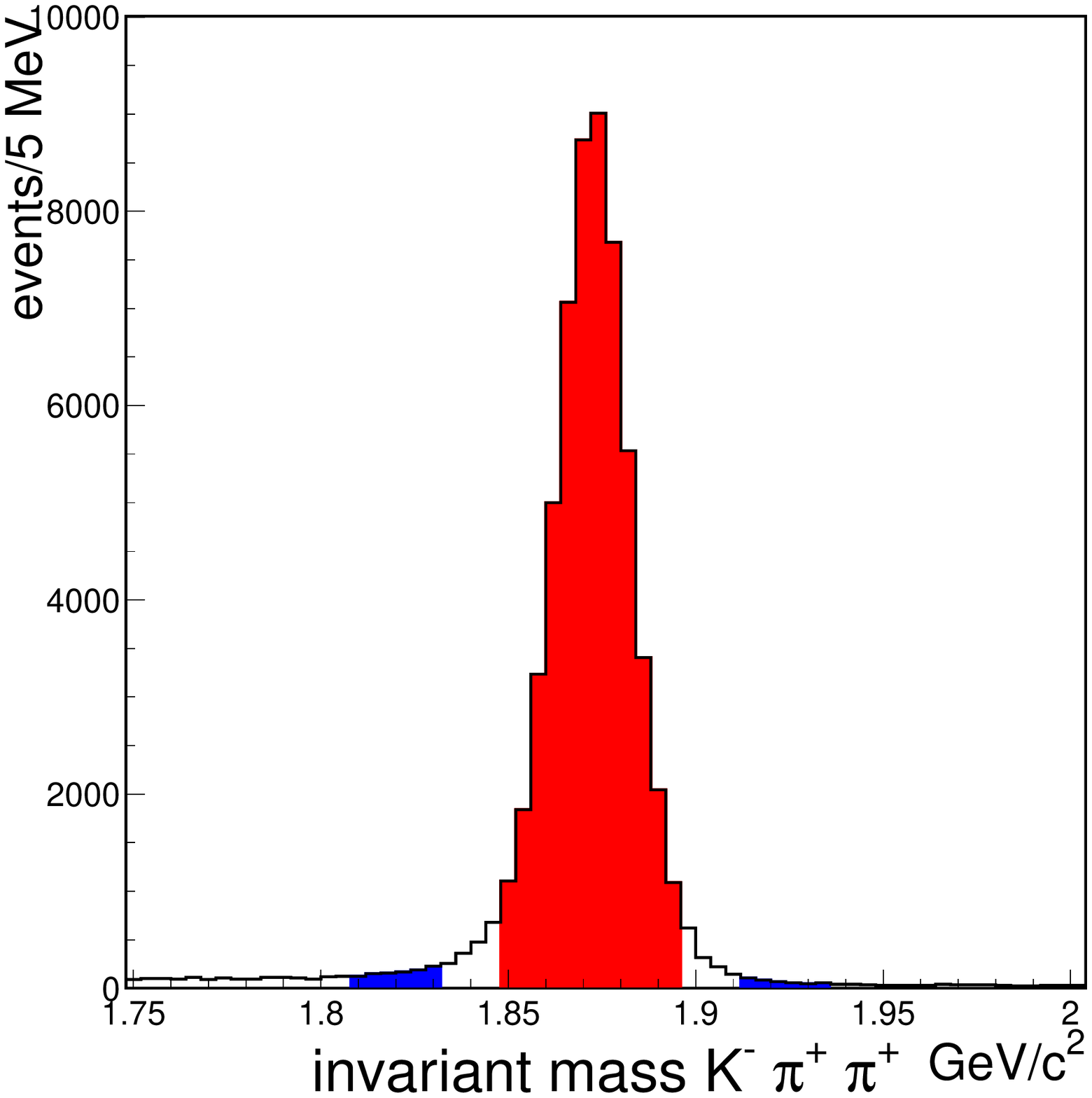}
}
 \caption{The $D^+ \to K^-\pi^+\pi^+$ Dalitz plot (left) and mass distribution (right): signal and sideband regions
 are indicated in red and blue respectively. The sidebands are at $\pm$(6--8)
 $\sigma$ from the peak.}
\label{signal}
\end{figure}

The recent FOCUS  study of the $D^+ \to K^-\pi^+\pi^+$ channel uses 53653
Dalitz-plot events with a signal fraction of $\sim$ 97\%, and represents the
highest statistics, most complete Dalitz plot analysis for this channel.
Invariant mass and Dalitz plot are shown in Fig.\ref{signal}.

Details of the analysis are in \cite{Focus_kpp}.

An additional complication in the $K\pi$ system comes from the presence in the
$S$-wave of the two isospin states, $I=1/2$ and $I=3/2$. Although only the
$I=1/2$ is dominated by resonances, both isospin components are involved in
the decay of the $D^+$ meson into $K^-\pi^+\pi^+$. A model for the decay
amplitudes of the two isospin states can be constructed from the 2 $\times$ 2
\emph{K-matrix} describing the $I=1/2$ $S$-wave scattering in $(K\pi)_1$ and
$(K\eta')_2$ (with the subscripts 1 and 2, respectively, labelling these two
channels), and
 the single-channel \emph{K-matrix} describing the $I=3/2$ $K^-\pi^+
\to K^-\pi^+$ scattering.

The \emph{K-matrix} form we use as input describes the $S$-wave $K^-\pi^+ \to
K^-\pi^+$ scattering from the LASS experiment \cite{lass} for energy above 825
MeV and $K^-\pi^- \to K^- \pi^-$ scattering from Estabrooks \emph{et al.}
\cite{estabrook}. The \emph{K-matrix} form follows the extrapolation down to
the $K\pi$ threshold for both $I=1/2$ and $I=3/2$ $S$-wave components by the
dispersive analysis by B\"uttiker \emph{et al.} \cite{butt}, consistent with
Chiral Perturbation Theory \cite{cpt}. The complete form is given below in
Eqs.~(\ref{eqn_K12}-\ref{eqn_K32}) with the parameters listed in
Table~\ref{tab_kmat12_value} \cite{mike_priv_com}.

The total $D$-decay amplitude can be written as

\begin{equation}
\mathcal{M}= {(F_{1/2})}_1(s) + F_{3/2}(s) + \sum_j a_j~
e^{i\delta_j}~B(abc|r), \label{A_tot_kmat}
\end{equation}
\par\noindent
where $s=M^2(K\pi)$, ${(F_{1/2})}_1$ and $F_{3/2}$ represent the $I=1/2$ and
$I=3/2$ decay amplitudes in the $K\pi$ channel, $j$ runs over vector and
spin-2 tensor resonances~\footnote{Higher spin resonances have been tried in
the fit with both formalisms but found to be statistically insignificant.},
and $~B(abc|r)$ are Breit--Wigner forms. The $J>0$ resonances should, in
principle, be treated in the same \emph{K-matrix} formalism. However, the
contribution from the vector wave comes mainly from the $K^*(892)$ state,
which is  well separated from the higher mass $K^*(1410)$ and $K^*(1680)$, and
the contribution from the spin-2 wave comes from $K_2^*(1430)$ alone. Their
contributions are limited to small percentages, and, as a first approximation,
they can be reasonably described by a simple sum of Breit--Wigners. More
precise results would require a better treatment of the overlapping
$K^*(1410)$ and $K^*(1680)$ resonances as well. In accord with SU(3)
expectations, the coupling of the $K\pi$ system to $K\eta$ is supposed to be
suppressed. Indeed we find little evidence that it is required.  Thus
$F_{1/2}$ is actually a vector consisting of two components: the first
accounting for the description of the $K\pi$ channel, the second of the
$K\eta'$ channel: in fitting $D^+ \to K^-\pi^+\pi^+$ we need, of course, the
${(F_{1/2})}_1$ element. Its form is

\begin{equation}
(F_{1/2})_1= (I-iK_{1/2}\rho)_{1j}^{-1}(P_{1/2})_j,
 \label{F_12}
\end{equation}
\par\noindent
where $I$ is the identity matrix, $K_{1/2}$ is the \emph{K-matrix} for the
$I=1/2$ $S$-wave scattering in $K\pi$ and $K\eta'$, $\rho$ is the
corresponding phase-space matrix for the two channels \cite{chung} and
$(P_{1/2})_j$ is the production vector in the channel $j$.

The form for $F_{3/2}$ is
\begin{equation}
F_{3/2}= (I-iK_{3/2}\rho)^{-1}P_{3/2},
 \label{F_32}
\end{equation}
\par\noindent
where $K_{3/2}$ is the single-channel scalar function describing the $I=3/2$
\, \mbox{$K^-\pi^+ \to K^-\pi^+$} scattering, and $P_{3/2}$ is the production
function into $K\pi$.

\begin{figure}[h]
  \centering
  \includegraphics[width=40 mm]{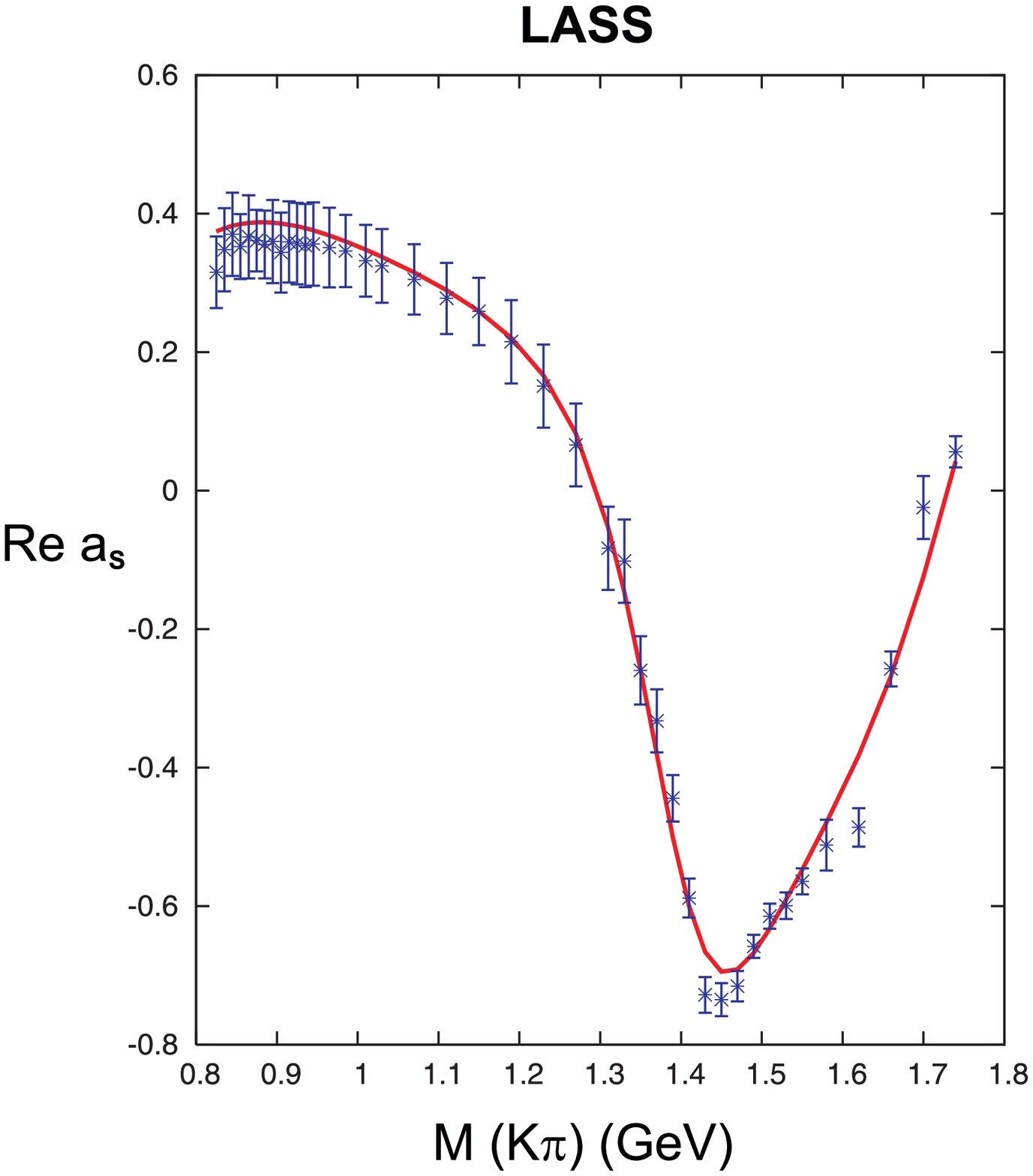}\hfil
  \includegraphics[width=40 mm]{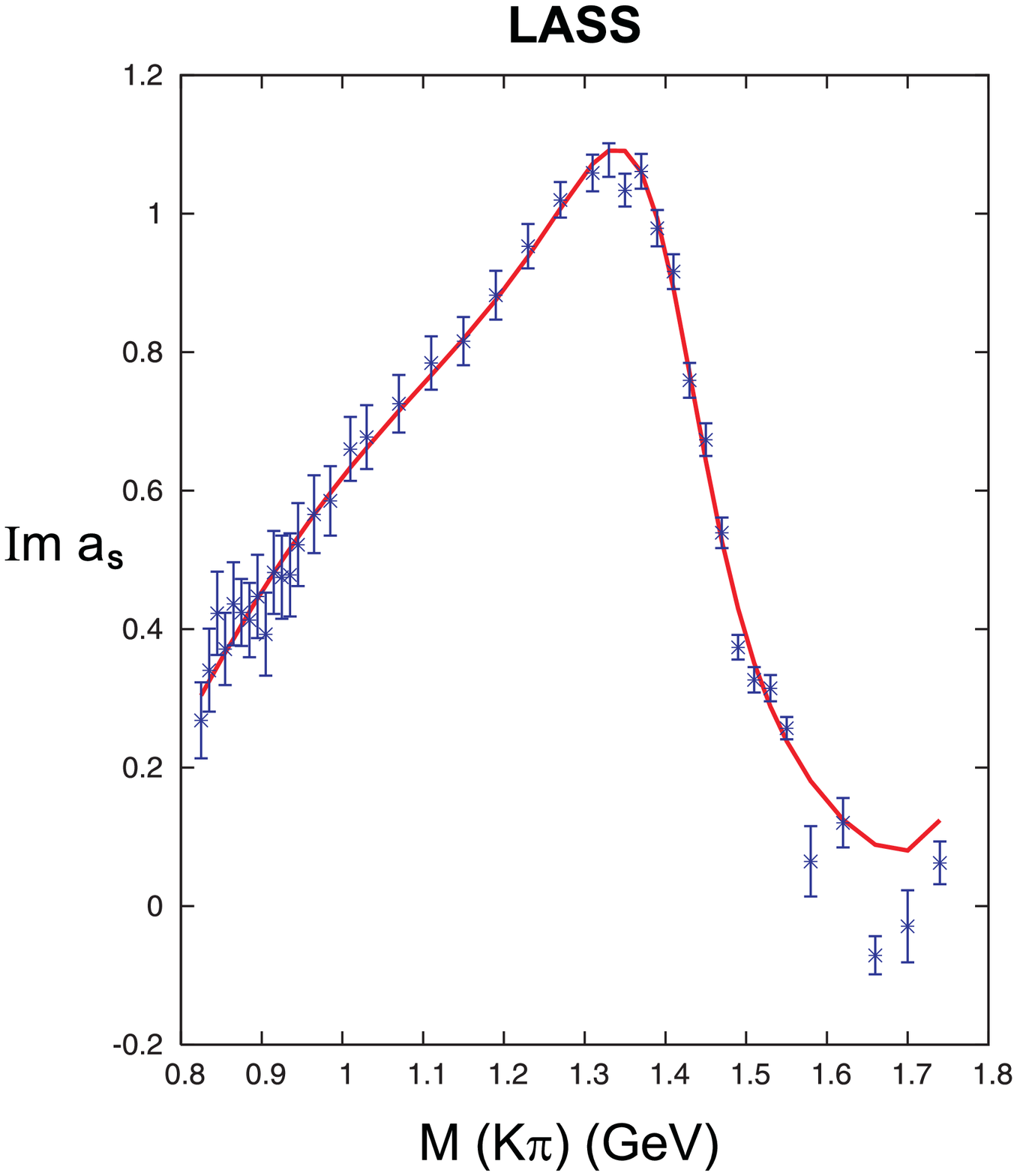}
  \caption{Real and imaginary $K^-\pi^+ \to K^-\pi^+$ amplitudes from the LASS experiment and their \emph{K-matrix} fit results. }
  \label{Lass_data}
\end{figure}

Fitting of the real and imaginary parts of the $K^-\pi^+ \to K^-\pi^+$ LASS
amplitude, shown in Fig.~\ref{Lass_data}, and using the predictions of Chiral
Perturbation Theory to continue this to threshold, gives the \emph{K-matrix}
parameters in Table~\ref{tab_kmat12_value}.

The $I=1/2$ \emph{K-matrix} is a single-pole, two-channel matrix whose
elements are given in Eq.~(\ref{eqn_K12}).

{\setlength\arraycolsep{2pt}
\begin{eqnarray}
 \label{eqn_K12}
K_{11} &=& \left(\!\frac{s{-}s_{0\frac{1}{2}}}{s_{\text{norm}}} \right)\!
\left(\frac{g_1 \cdot g_1}{s_1-s}+
C_{110} + C_{111}\tilde {s} + C_{112}\tilde {s}^2 \right) \nonumber \\
K_{22} &=&  \left(\!\frac{s{-}s_{0\frac{1}{2}}}{s_{\text{norm}}} \right)\!
\left(\frac{g_2 \cdot g_2}{s_1-s}+ C_{220} + C_{221}\tilde {s} + C_{222}\tilde {s}^2\right) \nonumber\\
K_{12} &=& \left(\!\frac{s{-}s_{0\frac{1}{2}}}{s_{\text{norm}}} \right)\!
\left(\frac{g_1 \cdot g_2}{s_1-s} + C_{120} + C_{121}\tilde {s} +
C_{122}\tilde {s}^2\right)\nonumber,
\\ & &
\end{eqnarray}

where the factor of $s_{norm} =  m_K^2+ m_{\pi}^2$ is conveniently introduced
to make the individual terms in the above expression dimensionless. $g_1$ and
$g_2$ are the real couplings of the $s_1$ pole to the first and the second
channel respectively. \mbox{$s_{0\frac{1}{2}}=0.23$ \, GeV$^2$} is the
position of the Adler zero in the $I=1/2$ ChPT elastic scattering amplitude
\footnote{ Chiral symmetry breaking demands an Adler zero in the elastic
$S$-wave amplitudes in the unphysical region. ChPT at next-to-leading order
fixes these positions $s_{0I}$ \cite{butt,cpt}.}. $C_{11i}$, $C_{22i}$ and
$C_{12i}$ for $i=0,1,2$ are the three coefficients of a second order
polynomial for the diagonal and off-diagonal elements of the symmetric
\emph{K-matrix}. Polynomials are expanded around $\tilde s = s/s_{norm} -1$.
This form generates an \emph{S-matrix} pole, which is conventionally quoted in
the complex energy plane as $E=M -i\Gamma/2=1.408 -{\it i} 0.110$ GeV. Any
more distant pole than $K_0^*(1430)$ is not reliably determined as this simple
\emph{K-matrix} expression does not have the required analyticity properties.
Nevertheless, it is an accurate description for real values of the energy,
where scattering takes place. Numerical values  of the terms in
Eq.~(\ref{eqn_K12}) are reported in Table~\ref{tab_kmat12_value}.

\begin{table*}[!htb]
 \begin{center}
  \caption{Values of parameters for the $I=1/2$  \emph{K-matrix}.}

   \begin{tabular}{|c|c|c|c|c|}
 \hline
 \textbf{pole (GeV\boldmath$^2$)} & \textbf{coupling (GeV)} & \boldmath{$C_{11i}$} & \boldmath{$C_{12i}$} & \boldmath{$C_{22i}$} \\
 \hline
 $ s_1=1.7919  $     &         &    &      & \\
          &   $ g_1= 0.31072$  &    &      & \\
          &   $ g_2=-0.02323$  &    &      & \\
          &          &        $C_{110}= 0.79299$ & $C_{120}= 0.15040$    &  $C_{220}=0.17054$     \\
          &          &        $C_{111}=-0.15099$ & $C_{121}= -0.038266$  &  $C_{221}=-0.0219$     \\
          &          &        $C_{112}= 0.00811$ & $C_{122}= 0.0022596$  &  $C_{222}=0.00085655$  \\
 \hline
  \end{tabular}
  \label{tab_kmat12_value}
  \end{center}
\end{table*}

The $I=3/2$ \emph{K-matrix} is given in Eq.~(\ref{eqn_K32}). Its form is
derived from a simultaneous fit to LASS data \cite{lass} and to $K^-\pi^- \to
K^-\pi^-$ scattering data \cite{estabrook}. It is a non-resonant, single-
channel scalar function.

\begin{equation}
\label{eqn_K32}
 K_{3/2} = \left(\frac{s-s_{0\frac{3}{2}}}{s_{\text{norm}}} \right)
\left( D_{110} + D_{111} \tilde {s} + D_{112} \tilde {s}^2 \right).
\end{equation}

\par\noindent
In Eq.~(\ref{eqn_K32}) $s_{0\frac{3}{2}}= 0.27$ \, GeV$^2$ is the Adler zero
position in the $I=3/2$ ChPT elastic scattering and the values of the
polynomial coefficients are \mbox{$D_{110} = -0.22147$}, \mbox{$D_{111}$ =
0.026637}, and \mbox{$D_{112} = -0.00092057$} \cite{mike_priv_com}.

When moving from scattering processes to $D$-decays, the production
\emph{P-vector} has to be introduced. While the \emph{K-matrix} is real,
\emph{P-vectors} are in general complex reflecting the fact that the initial
coupling $D^+\to (K^- \pi^+)\pi^+_{spectator}$ need not be real. The
\emph{P-vector} has to have the same poles as the \emph{K-matrix}, so that
these cancel in the physical decay amplitude. Their functional forms are:

\begin{equation}
(P_{1/2})_1= \frac{\beta g_1 e^{i\theta}} {s_1-s}
 + ( c_{10} + c_{11}\widehat{s}  + c_{12}
{\widehat{s}}^2 ) e^{i\gamma_1} \label{P_121}
\end{equation}

\begin{equation}
(P_{1/2})_2= \frac{\beta g_2 e^{i\theta}} {s_1-s} + (c_{20} + c_{21}\widehat
{s} +c_{22}{\widehat{s}}^2)e^{i\gamma_2} \label{P_122}
\end{equation}

\begin{equation}
P_{3/2}= (c_{30} + c_{31}\widehat{s} +c_{32}{\widehat{s}}^2)e^{i\gamma_3}
 \label{P_32}.
\end{equation}

\par\noindent
$\beta e^{i\theta} $ is the complex coupling to the pole in the `initial'
production process, $g_1$ and $g_2$ are the couplings as given by
Table~\ref{tab_kmat12_value}. The $K\pi$ mass squared \mbox{$s_c = 2$ \,
GeV$^2$} corresponds to the center of the Dalitz plot. It is convenient to
choose this as the value of $s$  about which the polynomials of
Eqs.~(\ref{P_121}-\ref{P_32}) are expanded, by defining $\widehat s = s -s_c$.
The polynomial terms in each channel are chosen to have a common phase
$\gamma_i$ to limit the number of free parameters in the fit and avoid
uncontrolled interference among the physical background terms. Thus, the
coefficients of the second order polynomial, $c_{ij}$, are real. Coefficients
and phases of the \emph{P-vectors}, except $g_1$ and $g_2$, are the only free
parameters of the fit determining the scalar components.

Free parameters for vectors and tensors are amplitudes and phases ($a_i$ and
$\delta_i$). $K\pi$ scattering determines the parameters of the {\it K-matrix}
elements and these are fixed inputs to this $D$ decay analysis.
Table~\ref{tab_P_vector_nov06} reports our \emph{K-matrix} fit results. It
shows quadratic terms in $(P_{1/2})_1$ are significant in fitting data, while
in both  $(P_{1/2})_2$ and $P_{3/2}$ constants are sufficient.
\begin{table*}[!htb]
 \begin{center}
  \caption{$S$-wave  parameters from the \emph{K-matrix} fit
 to the FOCUS $D^+ \to K^-\pi^+\pi^+$ data.  The first error is statistic, the second error is systematic from the experiment,
 and the third is systematic induced by model input parameters for higher resonances.
 Coefficients are for the unnormalized $S$-wave.}
 \label{tab_P_vector_nov06}
 \begin{tabular}{|c|c|}
 \hline
 \textbf{coefficient} & \textbf{phase (deg)}\\
 \hline
 $\beta$      = $3.389  \pm  0.152 \pm 0.002  \pm 0.068   $     & $\theta=    286 \pm 4 \pm 0.3 \pm 3.0   $  \\
 $c_{10}$     = $1.655  \pm  0.156 \pm 0.010  \pm 0.101   $     & $\gamma_1 = 304 \pm 6 \pm 0.4 \pm 5.8$ \\
 $c_{11}$     = $0.780  \pm  0.096 \pm 0.003  \pm 0.090   $     &      \\
 $c_{12}$     = $-0.954 \pm 0.058  \pm  0.0015 \pm 0.025  $     &      \\
 $c_{20}$     = $17.182 \pm 1.036  \pm 0.023 \pm 0.362    $     & $ \gamma_2 = 126  \pm  3 \pm 0.1 \pm 1.2  $ \\
 $c_{30}$     = $0.734  \pm 0.080  \pm 0.005 \pm 0.030    $     & $ \gamma_3 = 211  \pm 10 \pm 0.7\pm 7.8 $ \\
 \hline
 \multicolumn{2}{|c|}{
\emph{Total $S$-wave fit fraction}  = $ 83.23 \pm 1.50 \pm 0.04 \pm 0.07$ \%}  \\
 \multicolumn{2}{|c|}{
\emph{Isospin 1/2 fraction }  = $207.25 \pm 25.45 \pm  1.81 \pm 12.23$ \% } \\
 \multicolumn{2}{|c|}{
\emph{Isospin 3/2 fraction }  = $40.50  \pm 9.63 \pm 0.55 \pm 3.15$ \% } \\
 \hline
 \end{tabular}
 \end{center}
\end{table*}

The $J>0$ states required by the fit are listed in Table~\ref{tab_high_spin}.

\begin{table*}[!htb]
 \begin{center}
 \caption{Fit fractions, phases, and coefficients for the $J>0$ components from the \emph{K-matrix} fit
 to the FOCUS  $D^+ \to K^-\pi^+\pi^+$ data.  The first error is statistic, the
 second error is systematic from the experiment, and the
 third error is systematic induced by model input parameters for higher resonances. }
 \label{tab_high_spin}
 \begin{tabular}{|c|c|c|c|}
 \hline
 \textbf{component} & \textbf{fit fraction (\%)} & \textbf{phase \boldmath$\delta_j$ (deg)}
 & \textbf{coefficient} \\
 \hline
 $K^*(892)\pi^+$     &  $13.61 \pm  0.98 $       &    0 (fixed)                          & 1 (fixed)          \\
                     &  $ \pm \ 0.01 \pm 0.30 $  &                                       &                    \\

 $K^*(1680)\pi^+$    &  $ 1.90 \pm  0.63  $      &  $1   \pm  7  $       & $0.373 \pm 0.067 $ \\
                     &  $ \pm \ 0.009 \pm 0.43$  &  $ \pm \ 0.1 \pm 6 $  & $ \pm \ 0.009 \pm 0.047$ \\
 $K^*_2(1430)\pi^+$  &  $ 0.39 \pm  0.09  $      &  $296 \pm 7   $       & $0.169 \pm 0.017$ \\
                     &  $\pm \ 0.004 \pm 0.05 $  & $\pm\  0.3 \pm 1 $    & $ \pm \ 0.010 \pm 0.012 $ \\
 $K^*(1410)\pi^+$    &  $ 0.48 \pm  0.21  $      &  $293 \pm 17  $       & $0.188 \pm 0.041 $ \\
                     &  $ \pm \ 0.012 \pm 0.17$  & $\pm \ 0.4 \pm 7 $    & $ \pm \ 0.002 \pm 0.030$ \\
 \hline
 \end{tabular}
  \end{center}
\end{table*}
The $S$-wave component accounts for the dominant portion of the decay $(83.23
\pm 1.50) \%$. A significant fraction, $13.61\pm 0.98$\%, comes, as expected,
from $K^*(892)$; smaller contributions come from two vectors $K^*(1410)$ and
$K^*(1680)$ and from the tensor $K_2^*(1430)$.
It is conventional to quote fit fractions for each component and this is what
we do. Care should be taken in interpreting some of these since strong
interference can occur. This is particularly apparent between contributions in
the same-spin partial wave. While the total $S$-wave fraction is a sensitive
measure of its contribution to the Dalitz plot, the separate fit fractions for
$I=1/2$ and $I=3/2$ must be treated with care. The broad $I=1/2$ $S$-wave
component inevitably interferes strongly with the slowly varying $I=3/2$
$S$-wave, as seen for instance in \cite{mike_laura}. Fit results on the
projections are re shown in Fig.~\ref{fit_kmatrix_proj}. The corresponding
adaptive binning scheme is at the top of Fig.~\ref{fit_adapt} .

The fit $\chi^2$/d.o.f is 1.27 corresponding to a confidence level of 1.2\%.
If the $I=3/2$ component is removed from the fit, the $\chi^2$/d.o.f worsens
to 1.54, corresponding to a confidence level of $10^{-5}$.
\begin{figure}
 \centering
 \includegraphics[width=0.52\textwidth]{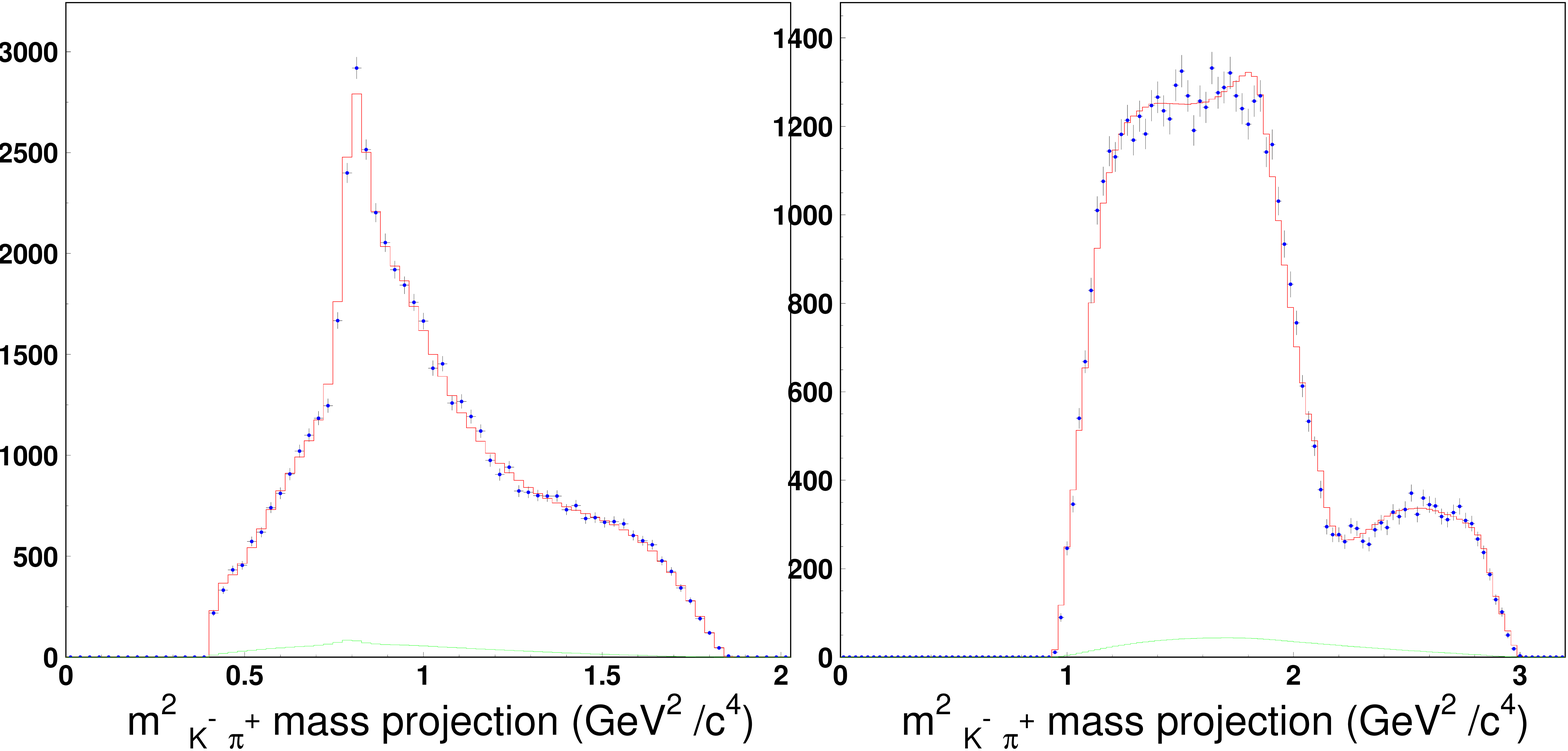}
 \caption{The Dalitz plot projections with the \emph{K-matrix} fit superimposed.
  The background shape under the signal is also shown.}
 \label{fit_kmatrix_proj}
%\end{center}
\end{figure}

\begin{figure}
 \centering
\includegraphics[width=68mm]{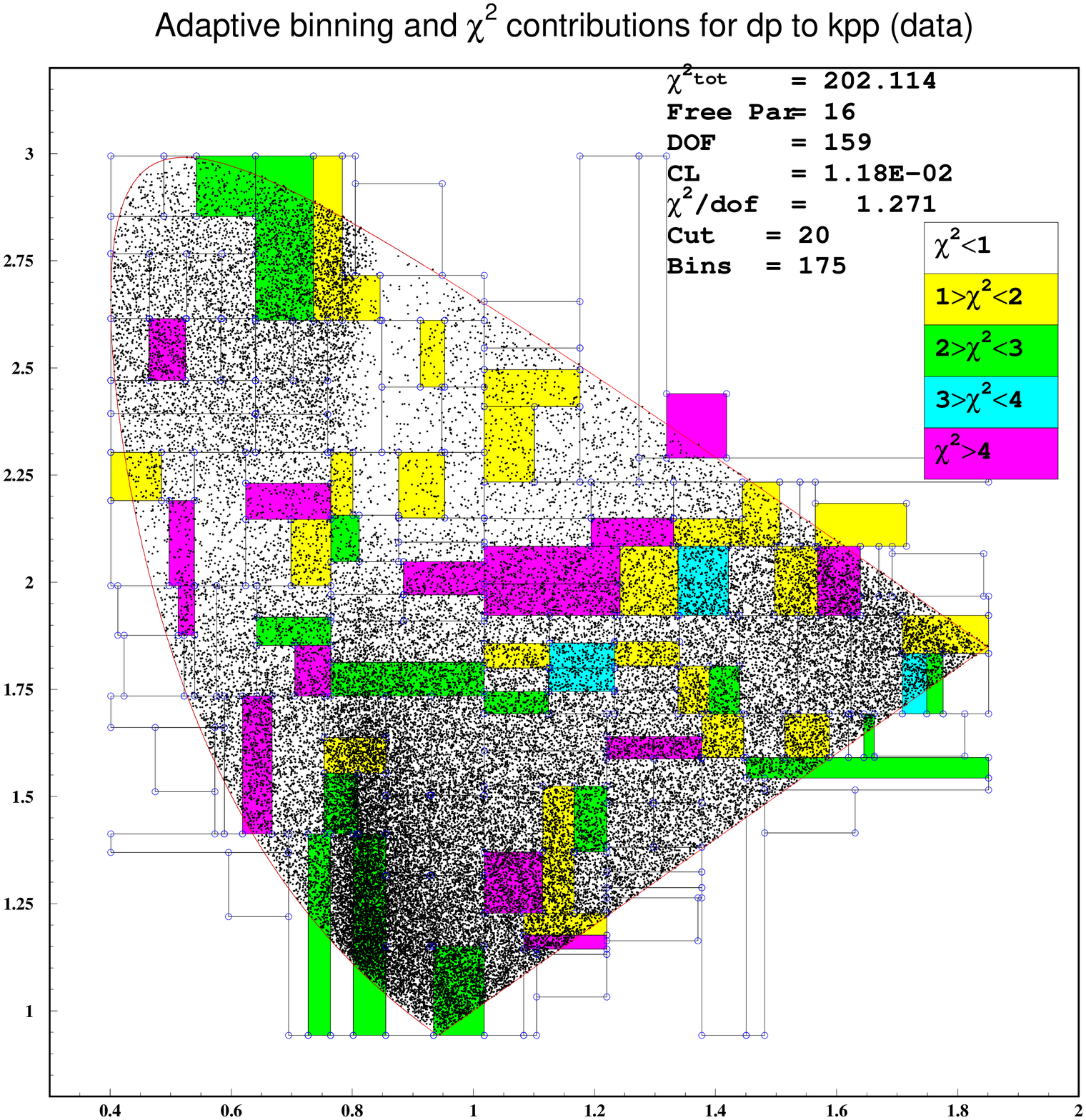}
\includegraphics[width=68mm]{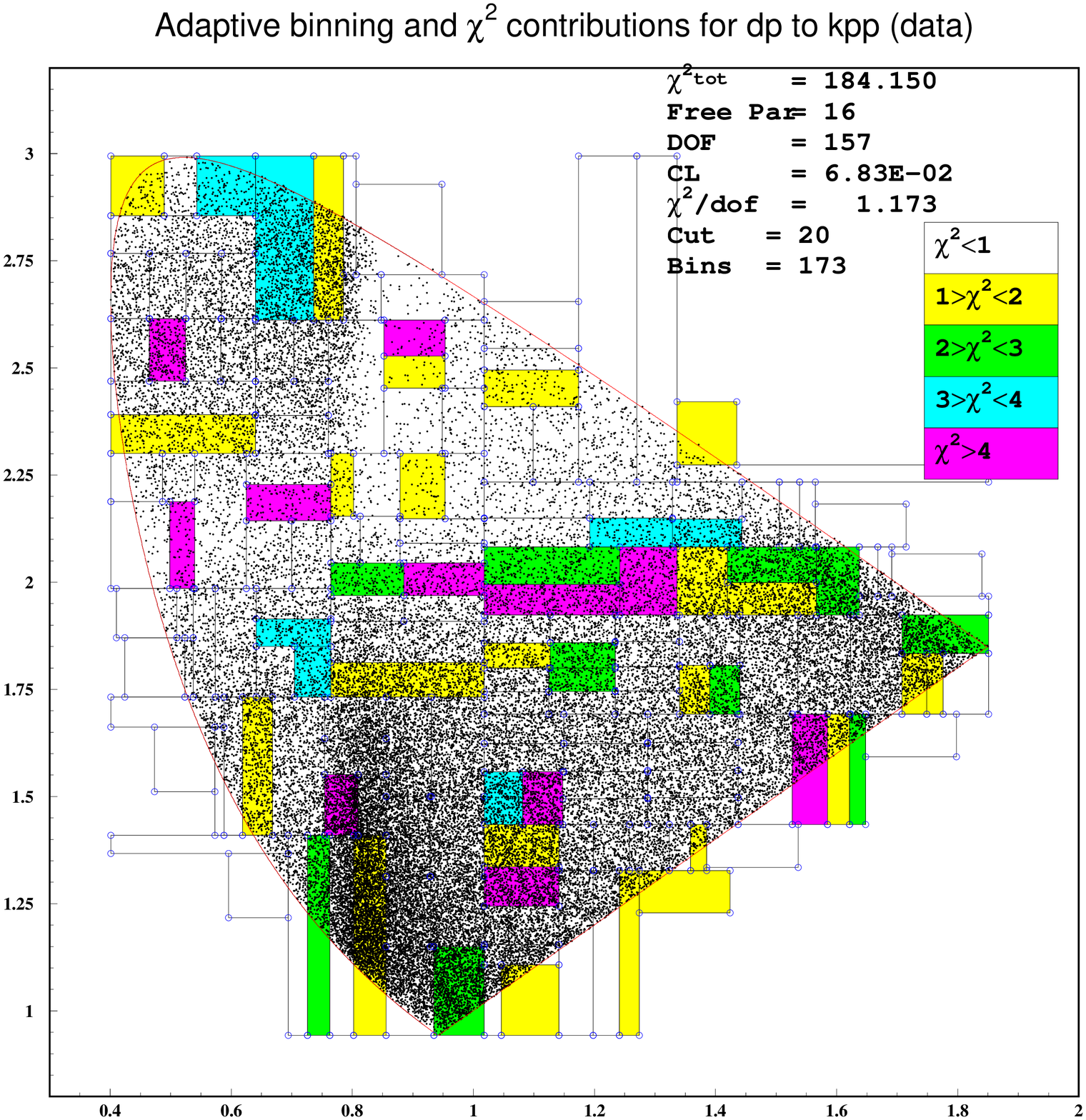}
\caption{ The adaptive binning schemes corresponding to the K-matrix (top) and
isobar (bottomt) fits. }
  \label{fit_adapt}
%\end{center}
\end{figure}

These results can be compared with those obtained in the effective isobar
model, which can serve as the standard for fit quality. Projections are shown
in Fig.~\ref{fit_isobar_results} and the adaptive binning scheme at the bottom
of Fig.~\ref{fit_adapt}.

\begin{figure}[h]
  \centering
  \includegraphics[width=0.52\textwidth]{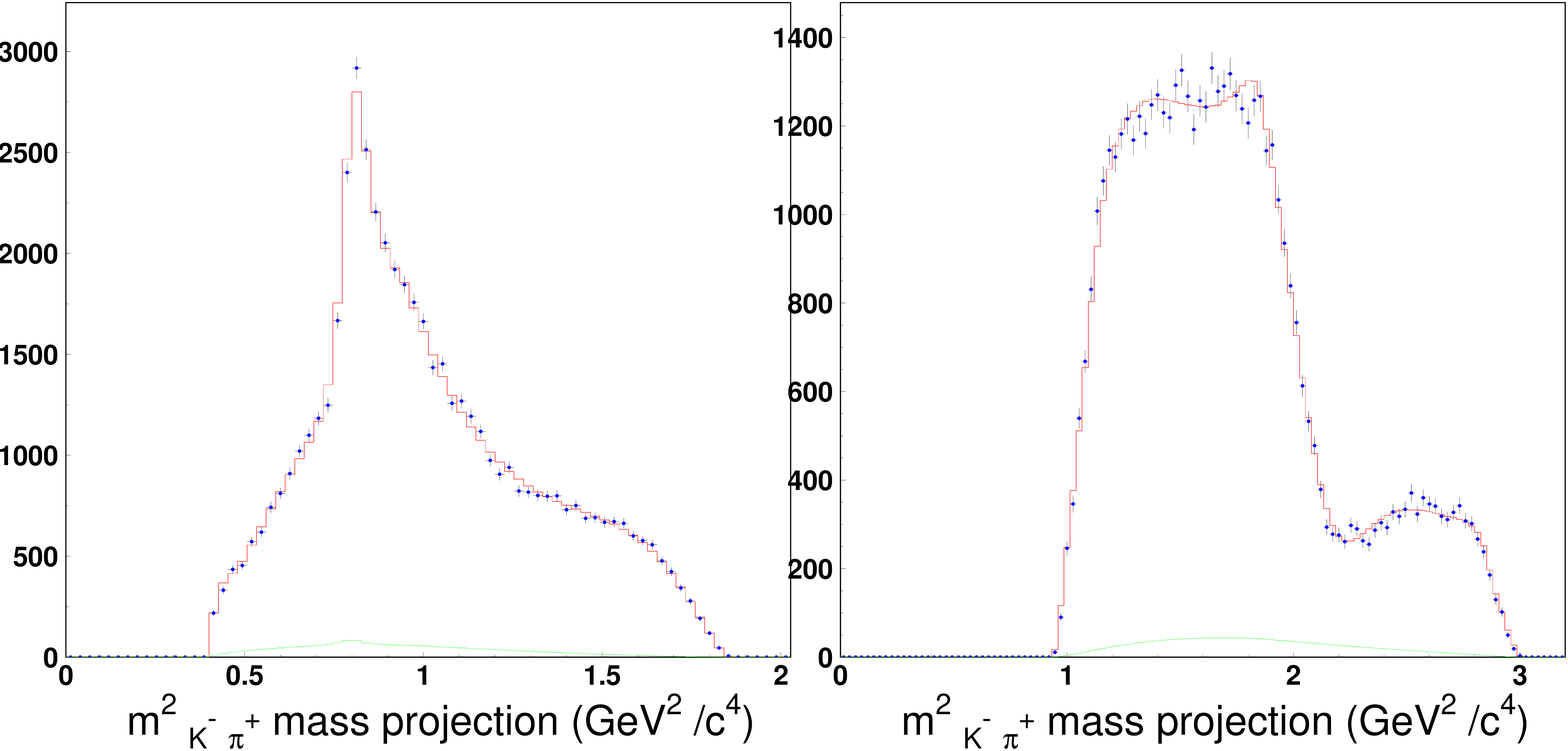}\\
  \caption{ Dalitz plot projections with our isobar fit superimposed.
  The background shape under the signal is also shown.}
  \label{fit_isobar_results}
%\end{center}
\end{figure}

Two \emph{ad hoc} scalar resonances are required, of mass $856 \pm 17$ and
$1461 \pm 4$ and width $464 \pm 28$ and $177 \pm 8$ MeV/$c^2$ respectively. A
detailed discussion of the results and the systematics can be found in
\cite{Focus_kpp}. The results of the \emph{K-matrix} fit showed that a
consistent representation with scattering is possible, the global fit quality
being indeed good. However, it deteriorates at higher $K\pi$ mass. This is not
surprising since our {\it K-matrix} treatment only includes two channels
$K\pi$ and $K\eta'$. While we have reliable information on the former channel,
we have relatively poor constraints on the latter. This means that as we
consider $K\pi$ masses far above $K\eta'$ threshold, these inadequacies in the
description of the $K\eta'$ channel become increasingly important. This is
expected to become worse as yet further inelastic channels open up.
Consequently, improvements could be made by using a number of $D$-decay chains
with $K\pi$ final state interactions and inputting all these in one combined
analysis in which several inelastic channels are included in the
\emph{K-matrix} formalism. In the present single $D^+\to K^-\pi^+\pi^+$
channel, adding further inelastic modes would be just adding free
unconstrained parameters for which there is little justification. It is
interesting to note that the adaptive binning scheme shows that both the
\emph{K-matrix} and the isobar fit are not able to reproduce data well in the
region at 2 \,GeV$^2$, in the vicinity of the $K\eta'$ threshold. It is also
the energy domain where higher spin states live. Vector and tensor fit
parameters in the two models are in very good agreement: we do not exclude the
possibility that a better treatment of these amplitudes could improve the
$\chi^2$. Some isolated spots of high $\chi^2$ could be caused by an imperfect
modeling of the efficiency as they are in the same regions in both fits.

A feature of the \emph{K-matrix} amplitude analysis is that it allows an
indirect phase measurement of the separate isospin components: it is this
phase variation with isospin $I=1/2$ which should be compared with the same
$I=1/2$ LASS phase, extrapolated from 825 GeV down to threshold according to
Chiral Perturbation Theory. This is done in the right plot of
Fig.~\ref{phase_tot_compa}. In this model \cite{aitch} the \emph{P-vector}
allows for a phase variation accounting for the interaction with the third
particle in the process of resonance formation. It so happens that the Dalitz
fit gives a nearly constant production phase. The two phases in
Fig.~\ref{phase_tot_compa}b) have the same behaviour up to $\sim$ 1.1 GeV.
However, approaching $K\eta'$ threshold, effects of inelasticity and differing
final state interactions start to appear.

% The inelastic channels in Kpi scattering and in Kpi interactions
%   in D decay will inevitably be different and we have no way of
%   including that without parametrizing these "higher" channels
The difference between the phases in Fig.~\ref{phase_tot_compa}a) is due to
the $I=3/2$ component.

\begin{figure}[h]
\centering
  \includegraphics[width=0.5\textwidth]{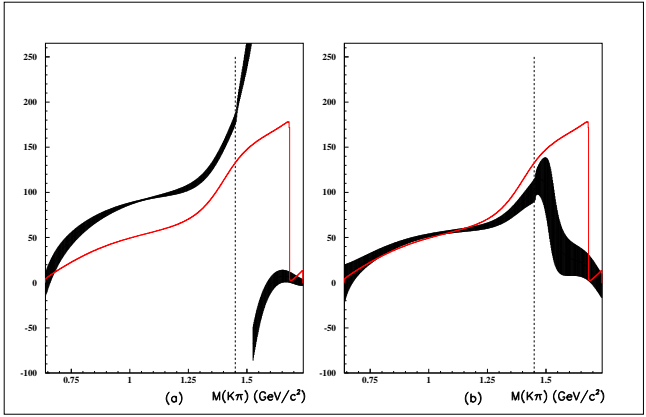}
  \caption{Comparison between the LASS $I=1/2$ phase + ChPT (continous line)
  and the \emph{F-vector} phases (with $\pm 1\,\sigma$ statical error bars); a)
   total \emph{F-vector} phase; b) $I=1/2$ \emph{F-vector} phase.
  The vertical dashed line shows the location of the $K\eta'$.}
  \label{phase_tot_compa}
%\end{center}
\end{figure}

These results are consistent with $K\pi$ scattering data, and consequently
with Watson's theorem predictions for two-body $K\pi$ interactions in the low
$K\pi$ mass region, up to $\sim$ 1.1 GeV, where elastic processes dominate.
This means that possible three-body interaction effects, not accounted for in
the \emph{K-matrix} parametrization, play a marginal role.

Our results for the total $S$-wave are in general agreement with those from
the E791 analysis, in which the $S$-wave modulus and phase were determined in
each $K\pi$ slice \cite{Brian}, \cite{Mike_china}.
What does this analysis contribute to the discussion of the existence and
parameters of the $\kappa$?  We know from analysis \cite{cherry} of the LASS
data (which in $K^-\pi^+$ scattering only start at 825 MeV) there is no pole,
the $\kappa(900)$, in its energy range. However, below 800 MeV, deep in the
complex plane, there is very likely such a state. Its precise location
requires a more sophisticated analytic continuation onto the unphysical sheet
than the {\it K-matrix} representation provided here. This is because of the
need to approach close to the crossed channel cut, which is not correctly
represented for a robust analytic continuation. However, our {\it K-matrix}
representation fits along the real energy axis inputs on scattering data and
Chiral Perturbation Theory in close agreement with those used in the analysis
by Descotes-Genon and Moussallam \cite{Desco_Mussalla} that locates the
$\kappa$ with a mass of $(658 \pm 13)$ MeV and a width of $(557 \pm 24)$ MeV
by careful continuation. These pole parameters  are quite different from those
implied by the simple isobar fits. We have thus shown that whatever $\kappa$
is revealed by our $D^+\to K^-\pi^+\pi^+$ results, it is the same as that
found in scattering data. Consequently, our analysis supports the conclusions
of \cite{Desco_Mussalla} and \cite{Zhou}.

\section{Conclusions}

Dalitz-plot analysis represents a unique, powerful and promising tool for
studying the Heavy Flavor decay dynamics. There is a recent, vigorous effort
to perform amplitude analysis: a more robust formalism has been implemented,
many channels have been investigated. The beauty community can benefit from
charm experience and expertise. The high statistic $D^+ \to K^-\pi^+\pi^+$
from FOCUS showed us that $D$-decay can also teach us about $K\pi$ interaction
much closer to threshold than the older scattering results. This serves as a
valuable check from experiment \cite{sanmike} of the inputs to the analyses of
\cite{Desco_Mussalla} and \cite{Zhou} based largely on theoretical
considerations. Dalitz-plot analysis will definitely keep us company over the
next few years. There will be a lot of work for both experimentalists and
theorists alike: synergy will be invaluable. Some complications have already
emerged, especially in the charm field, others, unexpected, will only become
clearer when we delve deeper into the beauty sector. $B_s$ will be a
completely new chapter. The analysis is challenging but there are no shortcuts
toward ambitious and high-precision studies and, ultimately, to New Physics
searches.

\end{document}